\documentclass[conference]{IEEEtran}
\usepackage{amsmath}
\usepackage{array}
\usepackage{amssymb,amsmath,color,graphicx,amsbsy,bm}
\usepackage{graphicx}
\usepackage{subfigure}
\usepackage{cite}
\usepackage{algorithm}
\usepackage{algorithmic}
\usepackage{amsmath}
\usepackage{amsfonts}
\usepackage{amssymb}
\usepackage{graphicx}
\usepackage{multirow}
\allowdisplaybreaks
\usepackage{soul}

\usepackage{color,soul}

\usepackage{textcomp}
\begin{document}
\IEEEoverridecommandlockouts
\IEEEpubid{\makebox[\columnwidth]{978-1-7281-8192-9/21/\$31.00 \copyright 2021 IEEE \hfill} \hspace{\columnsep}\makebox[\columnwidth]{ }}
%
% paper title
% Titles are generally capitalized except for words such as a, an, and, as,
% at, but, by, for, in, nor, of, on, or, the, to and up, which are usually
% not capitalized unless they are the first or last word of the title.
% Linebreaks \\ can be used within to get better formatting as desired.
% Do not put math or special symbols in the title.
\title{A Two-stage Stochastic Programming DSO Framework for Comprehensive Market Participation of DER Aggregators under Uncertainty}

% author names and affiliations
% use a multiple column layout for up to three different
% affiliations
\author{\IEEEauthorblockN{Mohammad Mousavi, \textit{Student Member, IEEE}}
\IEEEauthorblockA{Arizona State University\\
%Georgia Institute of Technology\\
%Atlanta, Georgia 30332--0250\\
Email: mmousav1@asu.edu}
\and
\IEEEauthorblockN{Meng Wu, \textit{Member, IEEE}}
\IEEEauthorblockA{Arizona State University\\
Email: mwu@asu.edu}
%\and
%\IEEEauthorblockN{James Kirk\\ and Montgomery Scott}
%\IEEEauthorblockA{Starfleet Academy\\
%San Francisco, California 96678--2391\\
%Telephone: (800) 555--1212\\
%Fax: (888) 555--1212}
}

% conference papers do not typically use \thanks and this command
% is locked out in conference mode. If really needed, such as for
% the acknowledgment of grants, issue a \IEEEoverridecommandlockouts
% after \documentclass

% for over three affiliations, or if they all won't fit within the width
% of the page, use this alternative format:
% 
%\author{\IEEEauthorblockN{Michael Shell\IEEEauthorrefmark{1},
%Homer Simpson\IEEEauthorrefmark{2},
%James Kirk\IEEEauthorrefmark{3}, 
%Montgomery Scott\IEEEauthorrefmark{3} and
%Eldon Tyrell\IEEEauthorrefmark{4}}
%\IEEEauthorblockA{\IEEEauthorrefmark{1}School of Electrical and Computer Engineering\\
%Georgia Institute of Technology,
%Atlanta, Georgia 30332--0250\\ Email: see http://www.michaelshell.org/contact.html}
%\IEEEauthorblockA{\IEEEauthorrefmark{2}Twentieth Century Fox, Springfield, USA\\
%Email: homer@thesimpsons.com}
%\IEEEauthorblockA{\IEEEauthorrefmark{3}Starfleet Academy, San Francisco, California 96678-2391\\
%Telephone: (800) 555--1212, Fax: (888) 555--1212}
%\IEEEauthorblockA{\IEEEauthorrefmark{4}Tyrell Inc., 123 Replicant Street, Los Angeles, California 90210--4321}}

% use for special paper notices
%\IEEEspecialpapernotice{(Invited Paper)}

% make the title area
\maketitle

\IEEEpubidadjcol
% As a general rule, do not put math, special symbols or citations
% in the abstract
\begin{abstract}
In this paper, a distribution system operator (DSO) framework is proposed for comprehensive retail and wholesale markets participation of distributed energy resource (DER) aggregators under uncertainty based on two-stage stochastic programming. Different kinds of DER aggregators including energy storage aggregators (ESAGs), demand response aggregators (DRAGs), electric vehicle (EV) aggregating charging stations (EVCSs), dispatchable distributed generation (DDG) aggregators (DDGAGs), and renewable energy aggregators (REAGs) are modeled. Distribution network operation constraints are considered using a linearized power flow. The problem is modeled using mixed-integer linear programming (MILP) which can be solved by using commercial solvers. Case studies are conducted to investigate the performance of the proposed DSO framework.

\end{abstract}

% no keywords

% For peer review papers, you can put extra information on the cover
% page as needed:
% \ifCLASSOPTIONpeerreview
% \begin{center} \bfseries EDICS Category: 3-BBND \end{center}
% \fi
%
% For peerreview papers, this IEEEtran command inserts a page break and
% creates the second title. It will be ignored for other modes.
\IEEEpeerreviewmaketitle

\section{Introduction}
% no \IEEEPARstart
The installed capacity of DERs is increasing, thanks to their low operational costs and growing demand. Being capable of providing fast ramping services, DER aggregators can effectively participate in the wholesale energy and regulation markets. However, uncontrolled participation of DER aggregators may cause security issues to distribution system operations. Hence, there is a need for an entity to enable DER aggregators to participate in the wholesale market and monitor the distribution system for secure and reliable operation.

Many topics have been examined in the context of market participation of DERs. In \cite{DiSomma,Baringo}, the concepts of DER aggregator and virtual power plant are introduced to enable DERs for wholesale market participation. A decentralized approach using Dantzig-Wolfe decomposition is presented for DER coordination in \cite{Anjos}. The proposed approach allows households to participate in the electricity market to minimize the total cost. In \cite{Liu,Lezama}, a microgrid is presented for wholesale market participation. The mentioned works ignore distribution grid operations. Hence, they neglect distribution grid security/reliability constraints which are necessary for DER's market participation. In \cite{Nguyen}, a biding strategy for market participation of a virtual power plant is proposed considering a demand response market which is considered as a stage between day-ahead and real-time markets. In \cite{Yang}, a bidding strategy is proposed for day-ahead and real-time markets participation of EV aggregators. In \cite{Nguyen,Yang}, in order to consider power balance equations, DC load flow is proposed, which is inappropriate due to high impedances in distribution grids. 

 Inspired by the smart grid technologies and growing DER installed capacity, the system operators call for a distribution level electricity market in which DERs can easily participate while assuring distribution grid security/reliability. The concept of distribution system operator (DSO) is presented recently in order to integrate DERs while operating the distribution network based on a retail market framework \cite{Faqiry,doPrado2,Parhizi}. In \cite{Faqiry}, a DSO is introduced for operating a day-ahead retail market. The distribution locational marginal price (D-LMP) is presented as a method for paying the market participants. However, the distribution network operation and corresponding security constraints are not included in the proposed model. In \cite{doPrado2}, the authors proposed a two-stage stochastic programming approach for a DSO to operate day-ahead energy and reserve markets. In \cite{Parhizi}, a distribution market operator (DMO) is proposed which collects offers from microgrids in order to participate in the wholesale market. To represent the relationship between D-LMP and transmission-level LMP, a penalty factor is defined. Both \cite{doPrado2} and \cite{Parhizi} adopt DC load flow, which is inappropriate for distribution grid modeling.

%In \cite{Bai}, a day-ahead market model operated by a DSO is presented. To determine the involvement of each participant, D-LMPs are decomposed to their components. In order to ensure power balance, AC power flow is presented by using second-order cone programming model. However, the participation of DERs in the wholesale market is not considered.\\ 

To the best of our knowledge, the DSO framework for comprehensive market participation of DER aggregators under uncertainty in the retail market as well as wholesale energy and regulation markets has not been studied yet. In this paper, a two-stage stochastic programming DSO framework is proposed for comprehensive market participation of DER aggregators under uncertainty. Various DER aggregators, including Energy storage aggregators (ESAGs), demand response aggregators (DRAGs), electric vehicle (EV) aggregating charging stations (EVCSs), dispatchable distributed generation (DDG) aggregators (DDGAGs), and renewable energy aggregators (REAGs), are considered. The proposed DSO optimally coordinates these DER aggregators for their participations in the retail market and wholesale energy/regulation markets, while maintaining distribution grid security. Case studies verify the effectiveness of the proposed DSO framework.
  
\section{Two-stage Stochastic DSO Formulation}\label{two-stage}
In this paper, the DSO is defined as an entity which interacts with DER aggregators and end-user customers on one side and trades with the wholesale market on the other side. The DSO collects offers from various types of DER aggregators and runs the retail market as well as coordinates the offers for constructing an aggregated offer for participating in the wholesale energy and regulation markets which is operated by the independent system operator (ISO) whose pay-for-performance regulation market is considered \cite{CAISO,Reza}.

The wholesale electricity market involves two stages: the day-ahead stage and balancing stage. For instance, California ISO (CAISO), which is adopted here, is a two-settlement market consisting of day-ahead and real-time markets, which is used for adjusting balance between supply and demand \cite{CAISO}. Market participants can participate in the day-ahead market and correct their share by participating in the real-time market in the case that their production or consumption has changed. In practice, usually, there is a difference between the offer of a participant and its production or consumption, especially for renewable energy producers. Hence, participation in the real-time market is necessary for them.

One important characteristic of a DSO is being capable of handling uncertainties in the system operation. An appropriate method for a market operator to cover uncertainties is using two-stage stochastic programming \cite{zhao2013pricing}. In this method, in the objective function, expected operational costs, including costs related to the day-ahead operation and costs related to the compensating actions in the real-time, is minimized. In this model, here-and-now variables are decisions related to the day-ahead market and wait-and-see variables are decisions related to the real-time market. Day-ahead market prices usually can be predicted with high accuracy \cite{Sadeghi-Mobarakeh}. Hence, sources of uncertainties are inelastic loads, renewable energy aggregator production, and real-time prices. The two-stage stochastic programming introduced in \cite{morales2013integrating} is adopted here.

\subsection{Objective Function}	
The DSO minimizes the distribution grid's total operational cost, considering 1) costs of buying/selling energy and selling regulation services to the wholesale energy and regulation markets; 2) costs of paying DER aggregators for their retail market participations. The objective function of the proposed two-stage stochastic programming is given by (\ref{equ.1}).
%\begin{equation}\label{equ.1}
\begin{align}\label{equ.1}
min &\sum_{t \in T}[ P^{sub}_{t}\pi^{e}_{t}-r^{sub,up}_{t}\pi^{cap,up}_{t}-r^{sub,dn}_{t}\pi^{cap,dn}_{t} \notag\\
&-r^{sub,up}_{t}S^{up}_{t}\mu^{up}_{t}\pi^{mil,up}_{t}-r^{sub,dn}_{t}S^{dn}_{t}\mu^{dn}_{t}\pi^{mil,dn}_{t} \notag\\
&+\sum_{k \in\{K2,K4\}}P_{t,k}\pi^{e}_{t,k}-\sum_{k_{3} \in K_{3}}P_{t,k_{3}}\pi^{e}_{t,k_{3}} \notag\\
&+\sum_{k \in K}[ r^{up}_{t,k}\pi^{cap,up}_{t,k}+r^{dn}_{t,k}\pi^{cap,dn}_{t,k}\\
&+r^{up}_{t,k}S^{up}_{t}\mu^{up}_{t}\pi^{mil,up}_{t,k}+r^{dn}_{t,k}S^{dn}_{t}\mu^{dn}_{t}\pi^{mil,dn}_{t,k}] \notag\\
&-\sum_{k_{1} \in K_{1}}\sum_{a \in A} P_{a,t,k_{1}}\pi^{e}_{a,t,k_{1}} \notag\\
& +\sum_{w \in W}\Omega_{w}(P^{sub,b,rl}_{t,w}\pi^{e,b,rl}_{t,w}-P^{sub,s,rl}_{t,w}\pi^{e,s,rl}_{t,w})] \notag
\end{align}
%\end{equation}
where $t$ and $T$ are the index and set for the entire operating timespan; $k$ and $K=\{K_{1}, K_{2}, K_{3}, K_{4}\}$ are the index and set for all DER aggregators; $k_{1}$ ($K_{1}$), $k_{2}$ ($K_{2}$), $k_{3}$ ($K_{3}$), $k_{4}$ ($K_{5}$), and $a$ ($A$) are the indices (sets) for all DRAGs, ESAGs, EVCSs, DDGAGs, and demand blocks, respectively; $P_t^{sub},r_t^{sub,up}$, and $r_t^{sub,dn}$ are the DSO's aggregated quantity offers to the wholesale energy, regulation capacity-up and capacity-down markets, respectively; $\pi_t^{e},\pi_t^{cap,up}$ ($\pi_t^{cap,dn}$), and $\pi_t^{mil,up}$ ($\pi_t^{mil,dn}$) are the wholesale energy, regulation capacity-up (capacity-down), and regulation mileage-up (mileage-down) prices, respectively; $P_{t,k},r_{t,k}^{up}$ and $r_{t,k}^{dn}$ are the energy, regulation capacity-up and capacity-down quantity offers made by DER aggregator $k$ with corresponding prices $\pi_{t,k}^{e},\pi_{t,k}^{cap,up},\pi_{t,k}^{cap,dn}$, respectively; $\mu_t^{up}$ and $\mu_t^{dn}$ are historical scores for providing regulation mileage-up and mileage-down services; $S_t^{up}$ and $S_t^{dn}$ are the regulation mileage-up and mileage-down ratios (the expected mileage for $1 MW$ provided regulation capacity); $P_{a,t,k_{1}}$ and $\pi^{e}_{a,t,k_{1}}$ are the power consumption and the corresponding energy price at each demand block; $\Omega_{w}$ is the probability of scenario $w$; $P_{t,w}^{sub,b,rl}$ is amount of power purchased from the wholesale real-time market with corresponding price $\pi_{t,w}^{e,b,rl}$; $P^{sub,s,rl}_{t,w}$ is amount of power sold to the wholesale real-time market with price $\pi_{t,w}^{e,s,rl}$.

In (\ref{equ.1}), the wholesale energy market is modeled as a producer in the DSO, while the wholesale regulation market is modeled as a consumer in the DSO. Therefore, cost terms related to the energy and regulation markets are associated with the positive and negative signs, respectively. The DSO is modeled as a price taker in the wholesale energy and regulation markets.

%note  that for the energy, the wholesale market is viewed as producer. Hence, it appears with positive sign. However, for the regulation market, it is viewed as demand. Hence, it is included with negative sign. 
%In (\ref{equ.1}), the first five terms are related to the the trade between the DSO and wholesale energy and regulation markets. Note  that, for the energy, the wholesale market is viewed as producer. Hence, it appears with positive sign. However, for the regulation market, it is viewed as demand. Hence, it is included with negative sign. The next seven terms are related to the DER aggregators participation. The last two terms are related to the trade between the DSO and wholesale energy market in the second stage which denotes real-time stage. 

\subsection{Constraints for Demand Response Aggregators (DRAGs)}
The operating constraints for DRAGs are as follows:
\begin{align}
& \sum_{a \in A} P_{a,t,k_{1}}-r_{t,k_{1}}^{cap,dn} \ge 0; \hspace{3mm}\forall t\in T,\, \forall k_{1} \in K_{1}\label{equ.2}\\
& \sum_{a \in A} P_{a,t,k_{1}}+r_{t,k_{1}}^{cap,up}\le \sum_{a \in A} P_{a,k1}^{max};\hspace{3mm}\forall t\in T,\, \forall k_{1} \in K_{1} \label{equ.3}\\
& 0 \le P_{a,t,k_{1}} \le P_{a,k1}^{max};\hspace{3mm}\forall a \in A,\,\forall t\in T,\, \forall k_{1} \in K_{1}\label{equ.4}\\
& 0 \le r_{t,k_{1}}^{cap,up} \le r_{t,k_{1}}^{cap,up,max};\hspace{3mm}\forall t\in T,\, \forall k_{1} \in K_{1} \label{equ.5}\\
& 0 \le r_{t,k_{1}}^{cap,dn} \le r_{t,k_{1}}^{cap,dn,max};\hspace{3mm}\forall t\in T,\, \forall k_{1} \in K_{1} \label{equ.6}
\end{align}
where $P_{a,t,k{1}}^{max}$ is the maximum power consumption at each demand block; $r_{t,k{1}}^{cap,up,max}$ and $r_{t,k{1}}^{cap,dn,max}$ are the maximum allowed regulation capacity-up and capacity-down quantity offers, respectively.

Equations (\ref{equ.2})-(\ref{equ.3}) ensure the total power consumed by the DRAG for buying/selling energy and offering regulation service is less than the maximum power consumption across all demand blocks within the DRAG. Equation (\ref{equ.4}) limits the amount of power offered by each demand block to its maximum value. Equations (\ref{equ.5})-(\ref{equ.6}) limit the regulation capacity-up and capacity-down quantity offers to their maximum values.

\subsection{Constraints for Energy Storage Aggregators (ESAGs)}
The operating constraints for ESAGs are as follows:
\begin{align}
\begin{split}\label{equ.7}
&P_{t,k_{2}}=E_{t-1,k_{2}}-E_{t,k_{2}}+(1/\eta_{k_{2}}^{di})r_{t,k_{2}}^{cap,up}\mu_{t}^{up}\\
&\hspace{10mm}-(\eta_{k_{2}}^{ch})r_{t,k_{2}}^{cap,dn}\mu_{t}^{dn};\hspace{3mm}\forall t\in T,\, \forall k_{2} \in K_{2} 
\end{split}\\
& P_{t,k_{2}}=(1/\eta_{k_{2}}^{di})P_{t,k_{2}}^{di}-(\eta_{k_{2}}^{ch})P_{t,k_{2}}^{ch};\hspace{3mm}\forall t\in T,\, \forall k_{2} \in K_{2}\label{equ.8}\\
& r_{t,k_{2}}^{cap,up}=r_{t,k_{2}}^{cap,up,di}+r_{t,k_{2}}^{cap,dn,ch};\hspace{3mm}\forall t\in T,\, \forall k_{2} \in K_{2} \label{equ.9}\\
& r_{t,k_{2}}^{cap,dn}=r_{t,k_{2}}^{cap,dn,di}+r_{t,k_{2}}^{cap,up,ch};\hspace{3mm}\forall t\in T,\, \forall k_{2} \in K_{2} \label{equ.10}\\
& E_{k_{2}}^{min} \le E_{t,k_{2}} \le E_{k_{2}}^{max};\hspace{3mm}\forall t\in T,\, \forall k_{2}  \in K_{2}\label{equ.11}\\
& 0 \le P_{t,k_{2}}^{di} \le b_{t,k_{2}} DR_{k_{2}}^{max} ;\hspace{3mm}\forall t\in T,\, \forall k_{2} \in K_{2}\label{equ.12}\\
& 0 \le r_{t,k_{2}}^{cap,up,di} \le b_{t,k_{2}} DR_{k_{2}}^{max};\hspace{3mm}\forall t\in T,\, \forall k_{2}  \in K_{2}\label{equ.13}\\
& 0 \le r_{t,k_{2}}^{cap,dn,di} \le b_{t,k_{2}} DR_{k_{2}}^{max};\hspace{3mm}\forall t\in T,\, \forall k_{2} \in K_{2} \label{equ.14}\\
& 0 \le P_{t,k_{2}}^{ch} \le (1-b_{t,k_{2}})CR_{k_{2}}^{max};\hspace{3mm}\forall t\in T,\, \forall k_{2}  \in K_{2}\label{equ.15}\\
& 0 \le r_{t,k_{2}}^{cap,up,ch} \le (1-b_{t,k_{2}})CR_{k_{2}}^{max};\hspace{0mm}\forall t\in T, \forall k_{2}  \in K_{2}\label{equ.16}\\
& 0 \le r_{t,k_{2}}^{cap,dn,ch} \le (1-b_{t,k_{2}})CR_{k_{2}}^{max};\hspace{0mm}\forall t\in T, \forall k_{2} \in K_{2} \label{equ.17}\\
\begin{split}\label{equ.18}
& r_{t,k_{2}}^{cap,dn,di} \le P_{t,k_{2}}^{di} \le DR_{k_{2}}^{max}-r_{t,k_{2}}^{cap,up,di};\\
&\hspace{27mm}\forall t\in T,\, \forall k_{2} \in K_{2}
\end{split}\\
\begin{split}\label{equ.19}
& r_{t,k_{2}}^{cap,dn,ch} \le P_{t,k_{2}}^{ch} \le CR_{k_{2}}^{max}-r_{t,k_{2}}^{cap,up,ch};\\
&\hspace{27mm}\forall t\in T,\, \forall k_{2} \in K_{2}
\end{split}
\end{align}
where $E_{t,k_{2}}$ is the charging level; $P_{t,k_{2}}^{ch}$ ($P_{t,k_{2}}^{di}$) and $\eta_{k_{2}}^{ch}$ ($\eta_{k_{2}}^{di}$) are the charging (discharging) power and charging (discharging) efficiencies, respectively; $r_{t,k_{2}}^{cap,up,ch}$ ($r_{t,k_{2}}^{cap,dn,ch}$) and $r_{t,k_{2}}^{cap,up,di}$ ($r_{t,k_{2}}^{cap,dn,di}$) are the regulation capacity-up (capacity-down) offers in charging and discharging modes, respectively; ${CR}_{k_{2}}^{max}$ and ${DR}_{k_{2}}^{max}$ are the maximum charging and discharging rates, respectively; $b_{t,k_{2}}$ is a binary variable indicating the charging ($b_{t,k_{2}}=0$) and discharging ($b_{t,k_{2}}=1$) modes.

ESAG's power injection is given by (\ref{equ.7}). ESAG's quantity offers for energy and regulation capacity-up/down markets are decomposed into charging and discharging terms by (\ref{equ.8})-(\ref{equ.10}). The charge level of ESAGs is limited by (\ref{equ.11}). Equations (\ref{equ.12})-(\ref{equ.17}) assure that ESAG's offers to the energy and regulation capacity-up/down markets are lower than their maximum values. In (\ref{equ.18})-(\ref{equ.19}), the total power offered by ESAG to the energy and regulation capacity-up/down markets lies within the charging and discharging rates.

\subsection{Constraints for EV Charging Stations (EVCSs)}
EVCSs are modeled as EV charging aggregators and are assumed to have unidirectional power flow. Constraints related to the operation of EVCSs are as follows: 
\begin{align}
& 0\le P_{t,k_{3}}\le ER_{k_{3}}^{max}b_{k_{3}};\hspace{3mm}\forall t\in T^{'},\, \forall k_{3} \in K_{3} \label{equ.20_1}\\
&0\le  r_{t,k_{3}}^{cap,up} \le ERR_{k_{3}}^{max}b_{k_{3}};\hspace{3mm}\forall t\in T^{'},\, \forall k_{3} \in K_{3} \label{equ.20_2}\\
&0\le  r_{t,k_{3}}^{cap,dn} \le ERR_{k_{3}}^{max}b_{k_{3}};\hspace{3mm}\forall t\in T^{'},\, \forall k_{3} \in K_{3} \label{equ.20_3}\\
& P_{t,k_{3}}+r_{t,k_{3}}^{cap,up} \le ER_{k_{3}}^{max};\hspace{3mm}\forall t\in T^{'},\, \forall k_{3} \in K_{3} \label{equ.20}\\
&  P_{t,k_{3}}-r_{t,k_{3}}^{cap,dn} \ge0;\hspace{3mm}\forall t\in T^{'},\, \forall k_{3}  \in K_{3}\label{equ.21}\\
%& r_{t,k_{3}}^{cap,up} \ge 0 ;\hspace{3mm}\forall t\in T^{'},\, \forall k_{3} \in K_{3}\label{equ.22}\\
\begin{split}\label{equ.23}
& 0.9CL_{k_{3}}^{max}b_{k_{3}}  \le E_{k_{3}}^{int}b_{k_{3}}+\sum_{t\in T^{'}}[ P_{t,k_{3}}+r_{t,k_{3}}^{cap,up} \mu_{t}^{up}\\
&\hspace{10mm}-r_{t,k_{3}}^{cap,dn} \mu_{t}^{dn}]  \gamma_{k_{3}}^{ch} \le CL_{k_{3}}^{max}b_{k_{3}} ;\hspace{1mm}\forall k_{3} \in K_{3}
\end{split}
\end{align}
where $T^{'}\subseteq T$ is the set of hours when EVs are available at the charging station; $ER_{k_{3}}^{max}$ is the maximum charging rate; $ERR_{k_{3}}^{max}$ is the maximum allowed regulation capacity offers, $CL_{k_{3}}^{max}$ is the maximum charge level; $E_{k_{3}}^{int}$ is the initial charge level; $\gamma_{k_{3}}^{ch}$ is the charging efficiency; $b_{k_{3}}$ is a binary variable which enables the DSO not to allocate the minimum power to EVCSs when their offering price is low. 

In (\ref{equ.20_1})-(\ref{equ.20_3}), EVCS's offers to the energy and regulation capacity-up/down markets are limited by their corresponding maximum values. In (\ref{equ.20})-(\ref{equ.21}), the total power offered by EVCS to the energy and regulation capacity-up/down markets lies within the maximum charging rate. Equation (\ref{equ.23}) assures the charge level of EVs is full. 

\subsection{Constraints for Dispatchable DG Aggregators (DDGAGs)}

The operating constraints for DDGAGs are as follows:
\begin{align}
&P_{t,k_{4}}+r_{t,k_{4}}^{cap,up} \le P_{k_{4}}^{max};\hspace{3mm}\forall t\in T,\, \forall k_{4} \in K_{4} \label{equ.24}\\
&P_{t,k_{4}}-r_{t,k_{4}}^{cap,dn} \ge P_{k_{4}}^{min};\hspace{3mm}\forall t\in T,\, \forall k_{4} \in K_{4} \label{equ.25} \\
&0\le r_{t,k_{4}}^{cap,up} \le RU_{k_{4}} ;\hspace{3mm}\forall t\in T,\, \forall k_{4} \in K_{4}\label{equ.26}\\
&0\le r_{t,k_{4}}^{cap,dn} \le RD_{k_{4}} ;\hspace{3mm}\forall t\in T,\, \forall k_{4} \in K_{4}\label{equ.27} 
\end{align} 
where $P_{k_{4}}^{max}$ and $P_{k_{4}}^{min}$ are the maximum and minimum power generations, respectively; $RU_{k_{4}}$ and $RD_{k_{4}}$ are the maximum ramp-up and ramp-down rates, respectively.

In (\ref{equ.24})-(\ref{equ.25}), the total power offered by DDGAG to the energy and regulation capacity-up/down markets lie within the DDGAG's maximum and minimum power generations. In (\ref{equ.26})-(\ref{equ.27}), the DDGAG's regulation capacity-up/down offers are limited by its maximum ramp-up/down rates. 

\subsection{Distribution Power Flow Equations}
The linearized power flow equations are adopted from \cite{Baran}:
\begin{align}
\begin{split}\label{equ.52}
&\sum_{k_{1}\in K_{1}}\sum_{a\in A}H_{n,k_{1}}P_{a,t,k_{1}}+\sum_{k_{3}\in K_{3}}H_{n,k_{3}} P_{t,k_{3}}+P_{t,n}^{D}\\
&-\sum_{k_{2}\in K_{2}}H_{n,k_{2}} P_{t,k_{2}}-\sum_{k_{4}\in K_{4}}H_{n,k_{4}} P_{t,k_{4}}\\
&-\sum_{k_{5}\in K_{5}}H_{n,k_{5}} P_{t,k_{5}}+H_{n}^{sub}P_{t}^{sub}+\sum_{j\in J}Pl_{j,t} A_{j,n} =0;\\
&\hspace{3mm}\forall t\in T, \, \forall n \in N \\
\end{split}\\
\begin{split}\label{equ.53}
&\sum_{k_{1}\in K_{1}}\sum_{a \in A}H_{n,k_{1}} P_{a,t,k_{1}} tan\phi_{k_{1}}+Q_{t,n}^{D}\\
&-\sum_{k_{4}\in K_{4}}H_{n,k_{4}} P_{t,k_{4}} tan\phi_{k_{4}}\\
&+H_{n}^{sub}Q_{t}^{sub}+\sum_{j\in J}Ql_{j,t} A_{j,n} =0 ;\hspace{3mm}\forall t\in T, \, \forall n \in N 
\end{split}\\
\begin{split}\label{equ.54}
& V_{m,t}=V_{n,t}-(r_{j} Pl_{j,t}+x_{j} Ql_{j,t} );\\
&\forall t\in T,\,\forall m\in N,\hspace{1mm}\forall n \in N,\, C(m,n)=1 ,\, A(j,n)=1 \\
\end{split}\\
& V^{min} \le V_{n,t} \le V^{max} ;\hspace{3mm}\forall t\in T,\,\forall n \in N \label{equ.55}\\
& -Pl^{max} \le Pl_{j,t} \le Pl^{max};\hspace{3mm}\forall t\in T,\, \forall j\in J  \label{equ.56}\\
& -Ql^{max} \le Ql_{j,t} \le Ql^{max};\hspace{3mm}\forall t\in T,\, \forall j\in J  \label{equ.57}\\
\begin{split}\label{equ.58}
& r_{t}^{sub,up}=\sum_{k_{2}\in K_{2}}r_{t,k_{2}}^{cap,up}+\sum_{k_{4}\in K_{4}}r_{t,k_{4}}^{cap,up}\\
&+\sum_{k_{1}\in K_{1}}r_{t,k_{1}}^{cap,dn}+\sum_{k_{3}\in K_{3}}r_{t,k_{3}}^{cap,dn};\hspace{3mm}\forall t\in T \\
\end{split}\\
\begin{split}\label{equ.59}
& r_{t}^{sub,dn}=\sum_{k_{2}\in K_{2}}r_{t,k_{2}}^{cap,dn}+\sum_{k_{4}\in K_{4}}r_{t,k_{4}}^{cap,dn}\\
&+\sum_{k_{1}\in K_{1}}r_{t,k_{1}}^{cap,up}+\sum_{k_{3}\in K_{3}}r_{t,k_{3}}^{cap,up};\hspace{3mm}\forall t\in T  
\end{split}\\
\begin{split}\label{equ.60}
&P_{t,n,w}^{D}-P_{t,n}^{D}-H_{n}^{sub}(P_{t,w}^{sub,RT}-P^{sub,b,RT}_{t,w})\\
&-\sum_{k_{5}\in K_{5}}H_{n,k_{5}} (P_{t,k_{5},w}-P_{t,k_{5}}-P_{t,k_{5},w}^{spill})\\
&+\sum_{j\in J}Pl_{j,t,w} A_{j,n}-\sum_{j\in J}Pl_{j,t} A_{j,n} =0;\\
&\hspace{3mm}\forall t\in T, \, \forall n \in N, \, \forall w \in W \\
\end{split}\\
\begin{split}\label{equ.61}
&Q_{t,n,w}^{D}-Q_{t,n}^{D}-H_{n}^{sub}Q_{t,w}^{sub,RT}+\sum_{j\in J}Ql_{j,t,w} A_{j,n}\\
&-\sum_{j\in J}Ql_{j,t} A_{j,n} =0 ;\hspace{3mm}\forall t\in T, \, \forall n \in N, \, \forall w \in W 
\end{split}\\
\begin{split}\label{equ.62}
& V_{m,t,w}-V_{m,t}=V_{n,t,w}-V_{n,t}-(r_{j} Pl_{j,t,w}-r_{j}Pl_{j,t}\\
&+x_{j} Ql_{j,t,w}-x_{j} Ql_{j,t} );\forall t\in T,\,\forall m\in N,\hspace{1mm}\forall n \in N,\,\\
& C(m,n)=1 ,\, A(j,n)=1, \, \forall w \in W \\
\end{split}\\
& V^{min} \le V_{n,t,w} \le V^{max} ;\hspace{3mm}\forall t\in T,\,\forall n \in N, \, \forall w \in W \label{equ.63}\\
& -Pl^{max} \le Pl_{j,t,w} \le Pl^{max};\forall t\in T, \forall j\in J, \forall w \in W  \label{equ.64}\\
& -Ql^{max} \le Ql_{j,t,w} \le Ql^{max};\forall t\in T, \forall j\in J, \forall w \in W  \label{equ.65}\\
&  P^{sub,b,rl}_{t,w}, P^{sub,s,rl}_{t,w} \ge 0;\forall t\in T, \forall w \in W \label{equ.44} 
%\begin{split}\label{equ.66}
%& r_{t}^{sub,up}=\sum_{k_{2}\in K_{2}}r_{t,k_{2}}^{cap,up}+\sum_{k_{4}\in K_{4}}r_{t,k_{4}}^{cap,up}+\sum_{k_{1}\in K_{1}}r_{t,k_{1}}^{cap,dn}+\sum_{k_{3}\in K_{3}}r_{t,k_{3}}^{cap,dn};\hspace{3mm}\forall t\in T \\
%\end{split}\\
%\begin{split}\label{equ.67}
%& r_{t}^{sub,dn}=\sum_{k_{2}\in K_{2}}r_{t,k_{2}}^{cap,dn}+\sum_{k_{4}\in K_{4}}r_{t,k_{4}}^{cap,dn}+\sum_{k_{1}\in K_{1}}r_{t,k_{1}}^{cap,up}+\sum_{k_{3}\in K_{3}}r_{t,k_{3}}^{cap,up};\hspace{3mm}\forall t\in T  
%\end{split}
\end{align}
where $k_{5}$ ($K_{5}$) are the indices (sets) for all REAGs; $P_{t,k_{5},w}^{spill}$ is the power of REAGs curtailed in each scenario; $H_{n,k}$ is the mapping matrix of DER aggregator $k$ to bus $n$; $P_{t,n}^{D}$ and $Q_{t,n}^{D}$ are the inelastic active and reactive power loads at each node; $Pl_{j,t}$ and $Ql_{j,t}$ are the active and reactive power flow at branch $j$; $A_{j,n}$ is the incidence matrix of branches and buses; $\phi$ is the phase angle; $C_{m,n}$ is the connecting nodes matrix.

Equations (\ref{equ.52})-(\ref{equ.59}) are related to the power flow equations in the day-ahead stage. Specifically, active and reactive power flows are represented by (\ref{equ.52})-(\ref{equ.53}); voltage drop at each line is represented by (\ref{equ.54}) and is limited by (\ref{equ.55}); active and reactive power limits at each line are represented by (\ref{equ.56})-(\ref{equ.57}); DSO's aggregated offers for participating in the wholesale energy and regulation capacity-up/down markets are represented by (\ref{equ.58})-(\ref{equ.59}). Equations (\ref{equ.60})-(\ref{equ.65}) are related to adjustments in the real-time stage. Specifically, Equations (\ref{equ.60})-(\ref{equ.62}) are active power, reactive power, and voltage adjustments, respectively; Equations (\ref{equ.63})-(\ref{equ.65}) ensure that bus voltages, line active and reactive power flows lie within their limits in each scenario, respectively. Equation (\ref{equ.44}) restricts the sign of trading power in the real-time stage.
\section{Case Studies}
\begin{figure}
	\centering
	\includegraphics[width=0.93\columnwidth, height = 0.65in]{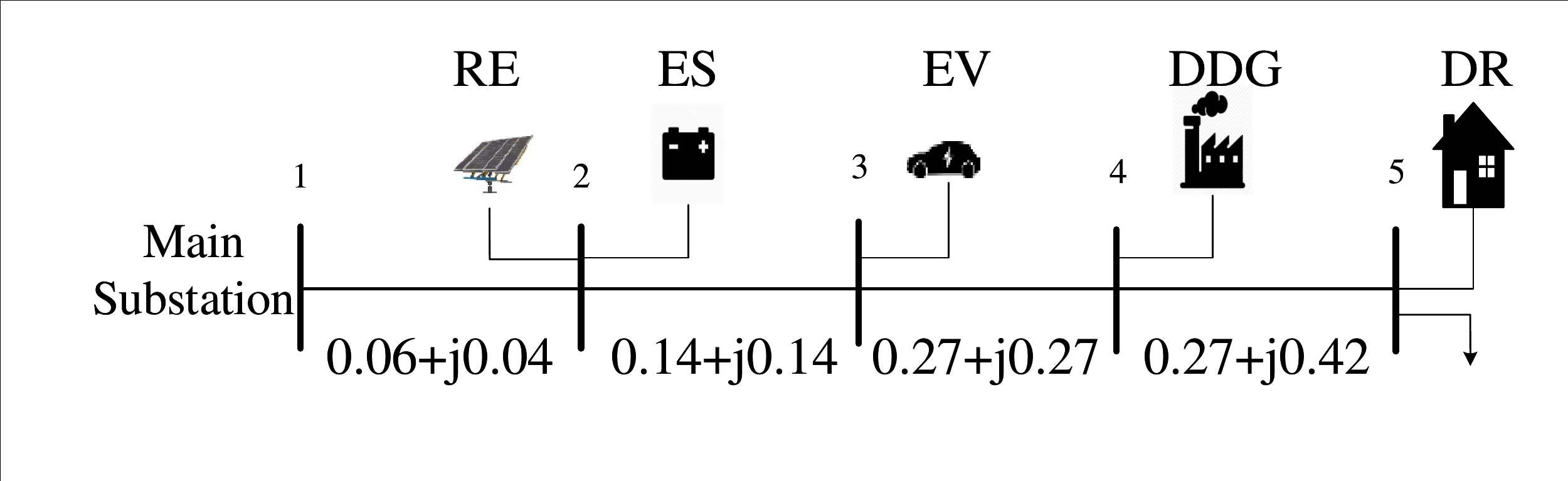}
	\caption{The small distribution network for case studies.}\label{fig.2.distribution network}
\end{figure}
In this section, two-stage stochastic programming introduced in Section \ref{two-stage} is used to obtain simulation results. Case studies are performed on the small distribution network in Fig.\ref{fig.2.distribution network}. The system contains $5$ nodes, where $N=\{1,2,3,4,5\}$; $4$ lines, where $J=\{1,2,3,4\}$; a DRAG, where $k_{1}=\{1\}$; an ESAG, where $k_{2}=\{2\}$; an EVCS, where $k_{3}=\{3\}$; a DDGAG, where $k_{4}=\{4\}$; a REAG, where $k_{5}=\{5\}$, and an inelastic load. The studies are performed over $24$ hours, $T=\{1,2,...,24\}$. EVs are available during Hours $16${\texttildelow}$24$, $T^{'}=\{16,17,...,24\}$. Initial charge level of ESAG is $8MW$. The following parameters are assumed: $\eta_{k_{2}}^{ch}=\eta_{k_{2}}^{di}=1$, $E_{k_{2}}^{min}=2MW$, $E_{k_{2}}^{max}=10MW$, $DR_{k_{2}}^{max}=CR_{k_{2}}^{max}=5MW$, $E_{k_{3}}^{int}=2MW$, $ER_{k_{3}}^{max}=5MW$, $ERR_{k_{3}}^{max}=0.5MW$, $P_{k_{4}}^{min}=0$,  $P_{k_{4}}^{max}=5MW$, $RU_{k_{4}}=RD_{k_{4}}=1MW$, $P_{a,t,k_{1}}^{max}=10MW$, $r_{k_{1}}^{cap,up,max}=r_{k_{1}}^{cap,dn,max}=1MW$.

In the deterministic case, inelastic load is considered to be $3$ $MW$ at all times and is located at Node $5$. Also, the maximum power production of REAG is considered to be $3$ $MW$. 
 Hourly energy prices, capacity up/down prices, and hourly regulation signals are generated by using hourly factors introduced in \cite{MM2} and are given in \cite{mousavi2020dso}. Case studies below focus on uncertainty. Market outcomes in deterministic cases can be found in \cite{mousavi2020dso}.

\subsubsection{Single Source of Uncertainty}
\begin{table}[!t]
	\centering
	\caption{REAG's production}\label{table.3 Renewable energy aggregator production}
	\begin{tabular}{c c c c c c}
		\hline
		Scenario Index & 1 & 2 & 3 & 4 &5 \\
		\hline
		Production (MW)&1&1.5&3&2&2.5\\
		Probability&0.1&0.1&0.6&0.1&0.1\\
		\hline
	\end{tabular}
\end{table}
 
\begin{figure}
	\centering
	\includegraphics[width=0.93\columnwidth, height = 1.2in]{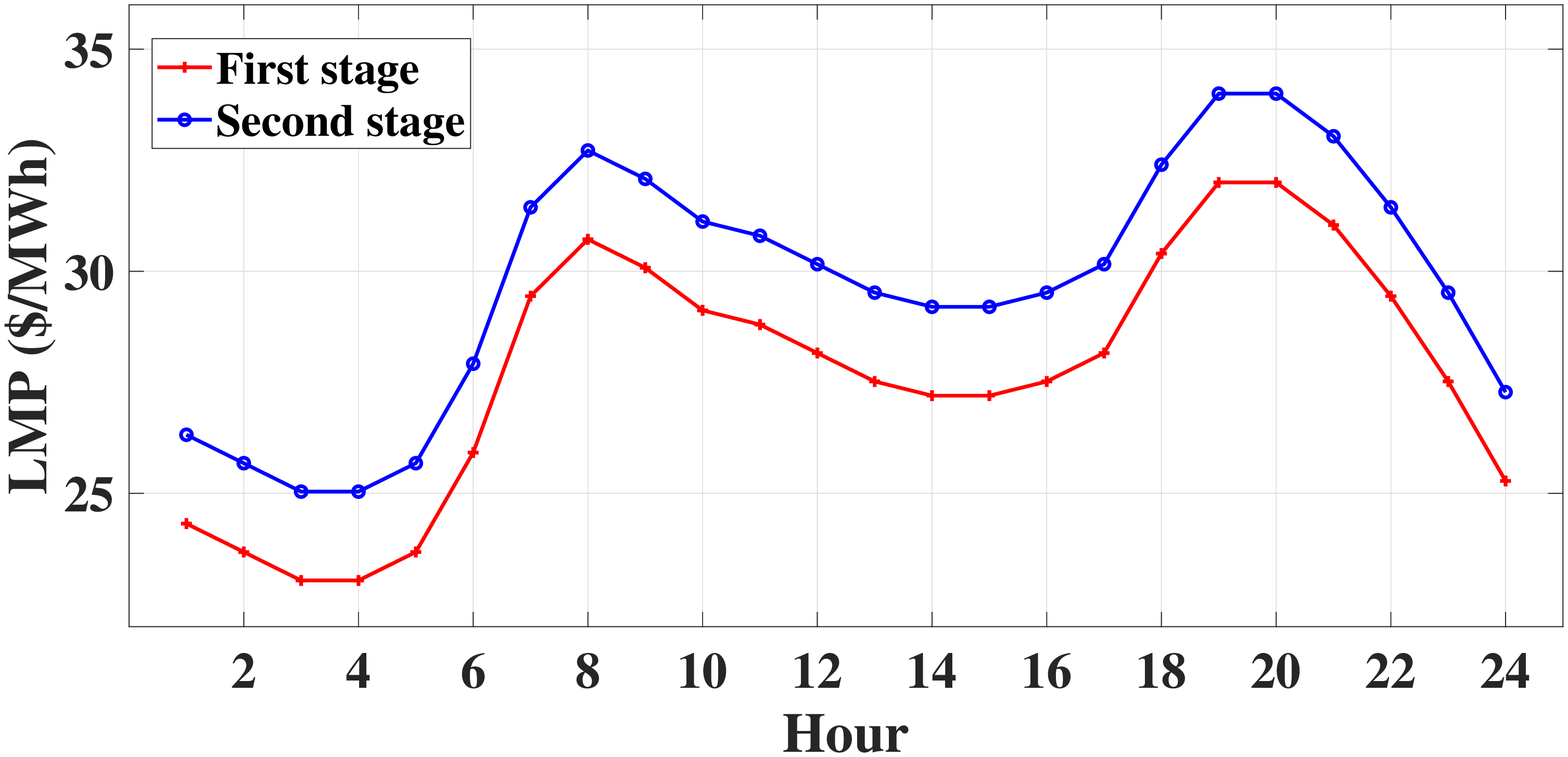}
	\caption{First-stage (day-ahead) and second-stage (real-time) LMPs under single source of uncertainty.}\label{fig.23.LMPtwostagestochastic}
\end{figure}

In this case, for simplicity, only one source of uncertainty is considered, which is the REAG production given in Table. \ref{table.3 Renewable energy aggregator production}. Wholesale real-time market prices are considered to be $2$ $\$/MWh$ higher than the corresponding day-ahead market prices. It is assumed the DSO can only buy energy from the real-time market. In two-stage stochastic programming, the first-stage LMP corresponds to the day-ahead market price, which is the dual variable of the power balance equation (\ref{equ.52}). The second-stage LMP corresponds to the real-time price, which is equal to the dual variable of power balance adjustment equation (\ref{equ.60}) divided by probability of occurrence of each scenario. Fig. \ref{fig.23.LMPtwostagestochastic} shows the first-stage (day-ahead) and second-stage (real-time) LMPs. Market participants are first settled by day-ahead LMPs. After that, market participants which need real-time compensation due to their uncertainties are settled by real-time LMPs.

\begin{figure}
	\centering
	
	\subfigure[]
	{
		\includegraphics[width=0.45\textwidth]{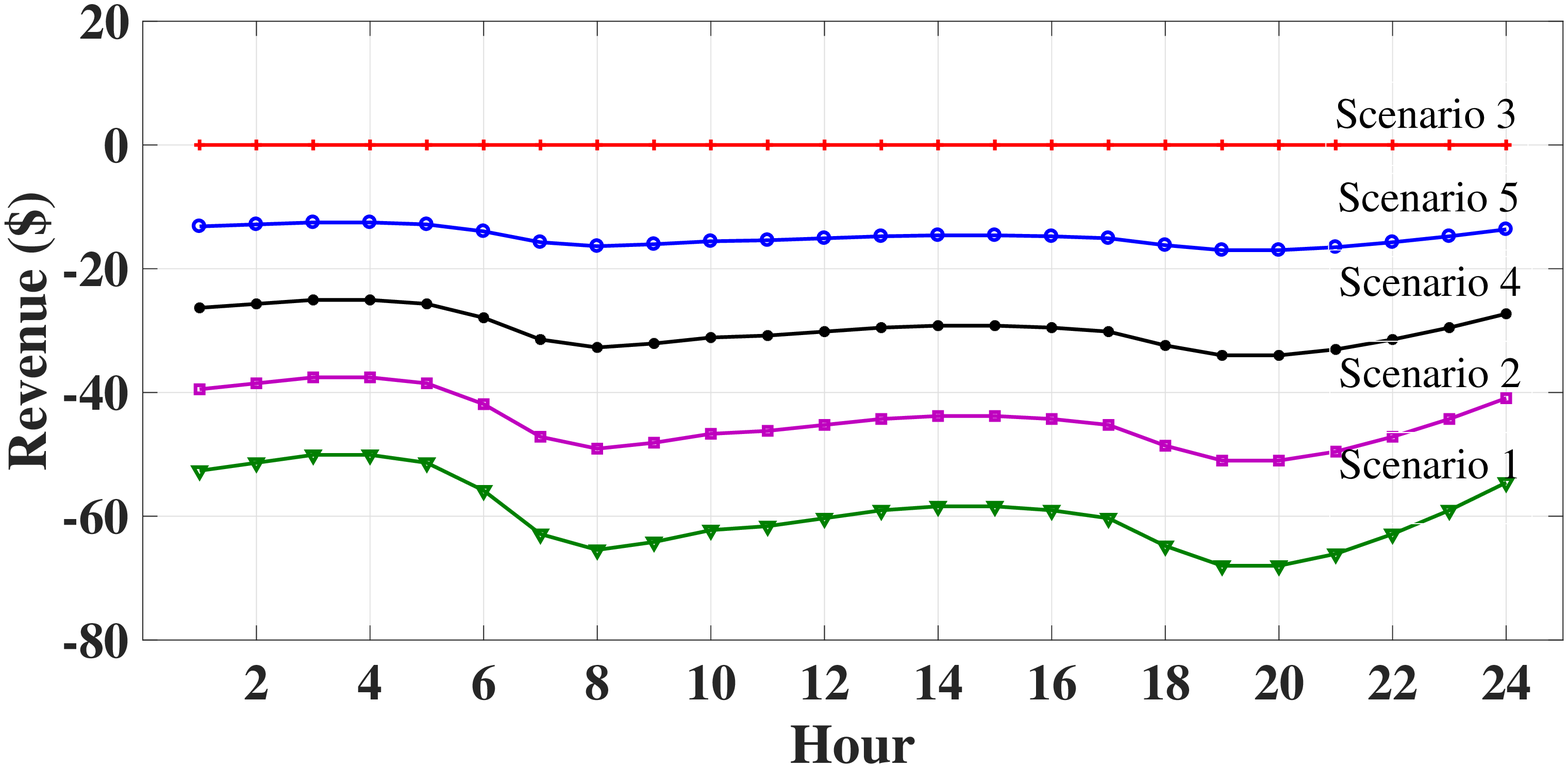}
		\label{fig.24.renewableaggregatorsecondstage}
	}
	\subfigure[]
	{
		\includegraphics[width=0.45\textwidth]{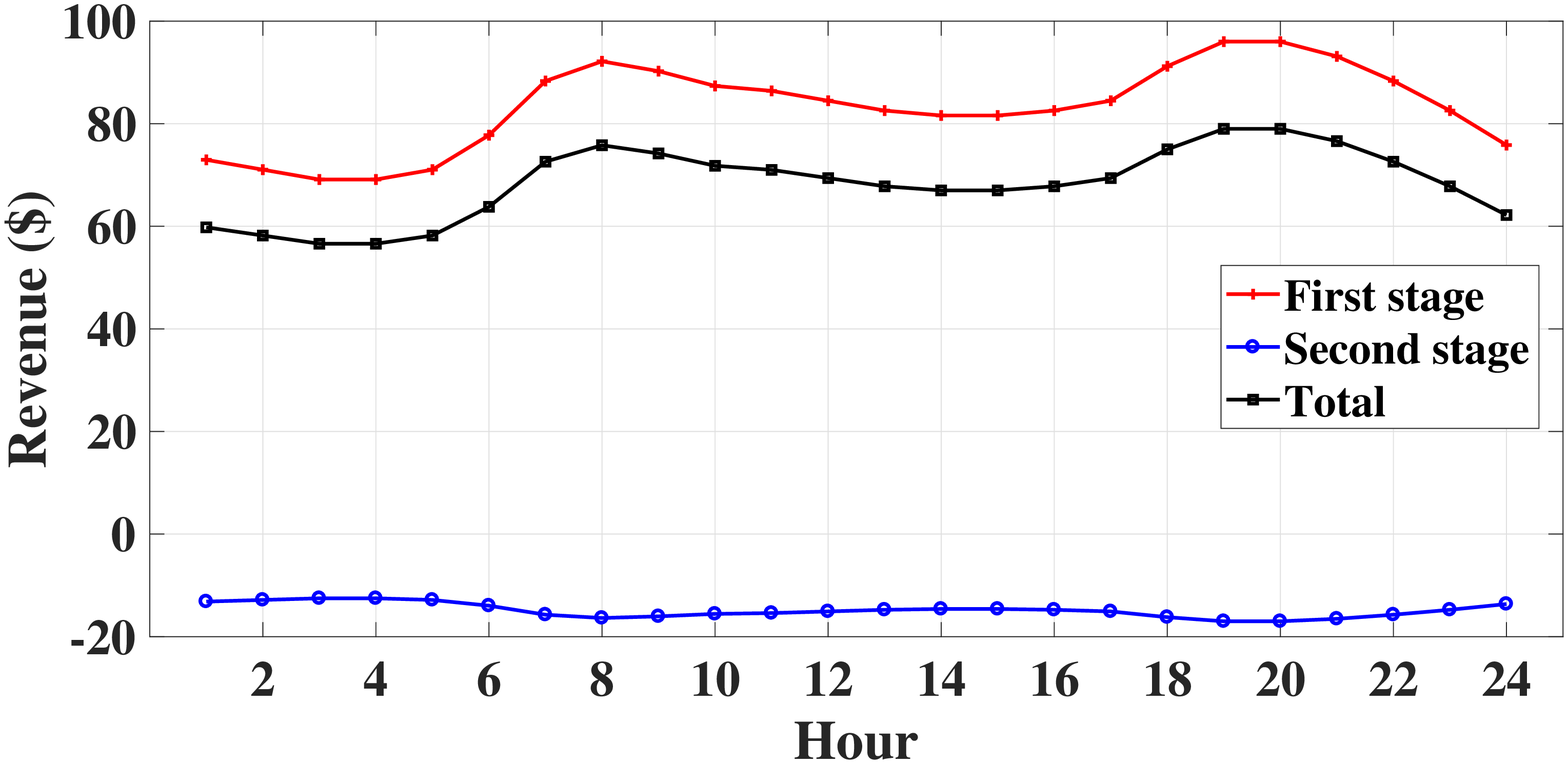}
		\label{fig.25.renewableaggregatorrevenue}
	}
	
	\caption{Under single source of uncertainty, (a) REAG's second-stage (real-time) revenue under each scenario; (b) REAG's first-stage (day-ahead) revenue, expected second-stage (real-time) revenue, and total expected revenue.}\label{Fig.26.renewableaggregatorrevenue}
\end{figure}

Fig. \ref{fig.24.renewableaggregatorsecondstage} shows the REAG's second-stage (real-time) revenue in each scenario. In Scenario 3, REAG's scheduled power in the day-ahead stage is the same as that in the real-time stage. Hence, there is no need for real-time correction. In other scenarios, REAG's scheduled power in the day-ahead stage is higher than that in the real-time stage. This power deficiency should be compensated by purchasing from the wholesale real-time market. As a result, the REAG's second-stage (real-time) revenue is negative, which means it purchases power from the wholesale real-time market. Fig. \ref{fig.25.renewableaggregatorrevenue} shows the REAG's first-stage (day-ahead) revenue, expected second-stage (real-time) revenue, and expected total revenue.

\begin{figure}
	\centering
	\includegraphics[width=0.93\columnwidth, height = 1.2in]{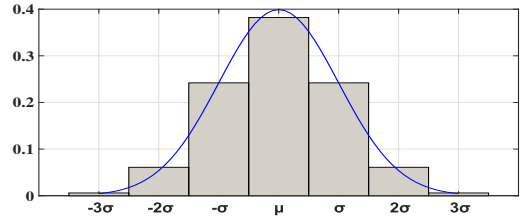}
	\caption{Normal distribution used under multiple sources of uncertainties.}\label{fig.30.normaldistribution}
\end{figure} 
 
\subsubsection{Multiple Sources of Uncertainties} As mentioned above, there are three sources of uncertainties including REAG production, inelastic load, and real-time prices. Random scenarios can be generated using scenario generation methods based on the probability distribution function. Scenario reduction methods can be applied to reduce computation burden. In this case, for simplicity, normal distribution in Fig. \ref{fig.30.normaldistribution} with mean value $\mu$ and standard deviation $\sigma$ is considered as the probability distribution of random variables. Seven scenarios from $-3\sigma$ to $3\sigma$ are considered. The mean value of each random variable is assumed to be the same as its value in the deterministic case. The standard deviation $\sigma$ is considered to be $5\%$, $15\%$, and $8\%$ for real-time prices, inelastic load, and REAG production, respectively. The REAG production scenarios are considered to change in the opposite direction of the real-time prices and inelastic load. In the second-stage (real-time), the price of selling energy to the wholesale market is considered to be 0.8 of the price of buying energy from it.

\begin{figure}
	\centering
	\includegraphics[width=0.93\columnwidth]{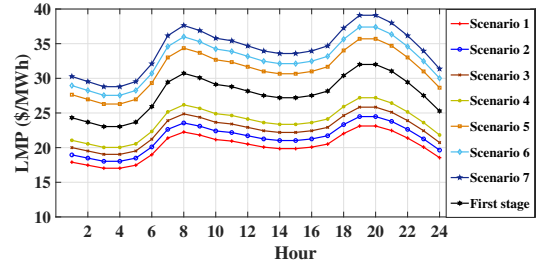}
	\caption{Under multiple sources of uncertainties, the REAG's first-stage (day-ahead) LMP and second-stage (real-time) LMPs in different scenarios.}\label{fig.31.LMPsecondstage3}
\end{figure}

Fig. \ref{fig.31.LMPsecondstage3} shows the first-stage (day-ahead) LMPs and second-stage (real-time) LMPs in different scenarios. LMPs in Scenarios $1${\texttildelow}$3$ equal the real-time prices of selling energy to the wholesale market, since in these scenarios, the demand is lower than the production in the retail market operated by the DSO. However, in Scenarios $5${\texttildelow}$7$ the LMPs equal the real-time prices of buying energy from the wholesale market, since in these scenarios the demand is greater than the production.  

\subsubsection{Sensitivity Analysis}
Sensitivity analysis is carried out on the REAG's revenue with respect to changing the real-time prices in both previous case studies. 

\begin{figure}
	\centering

\subfigure[]
{
	\includegraphics[width=0.93\columnwidth]{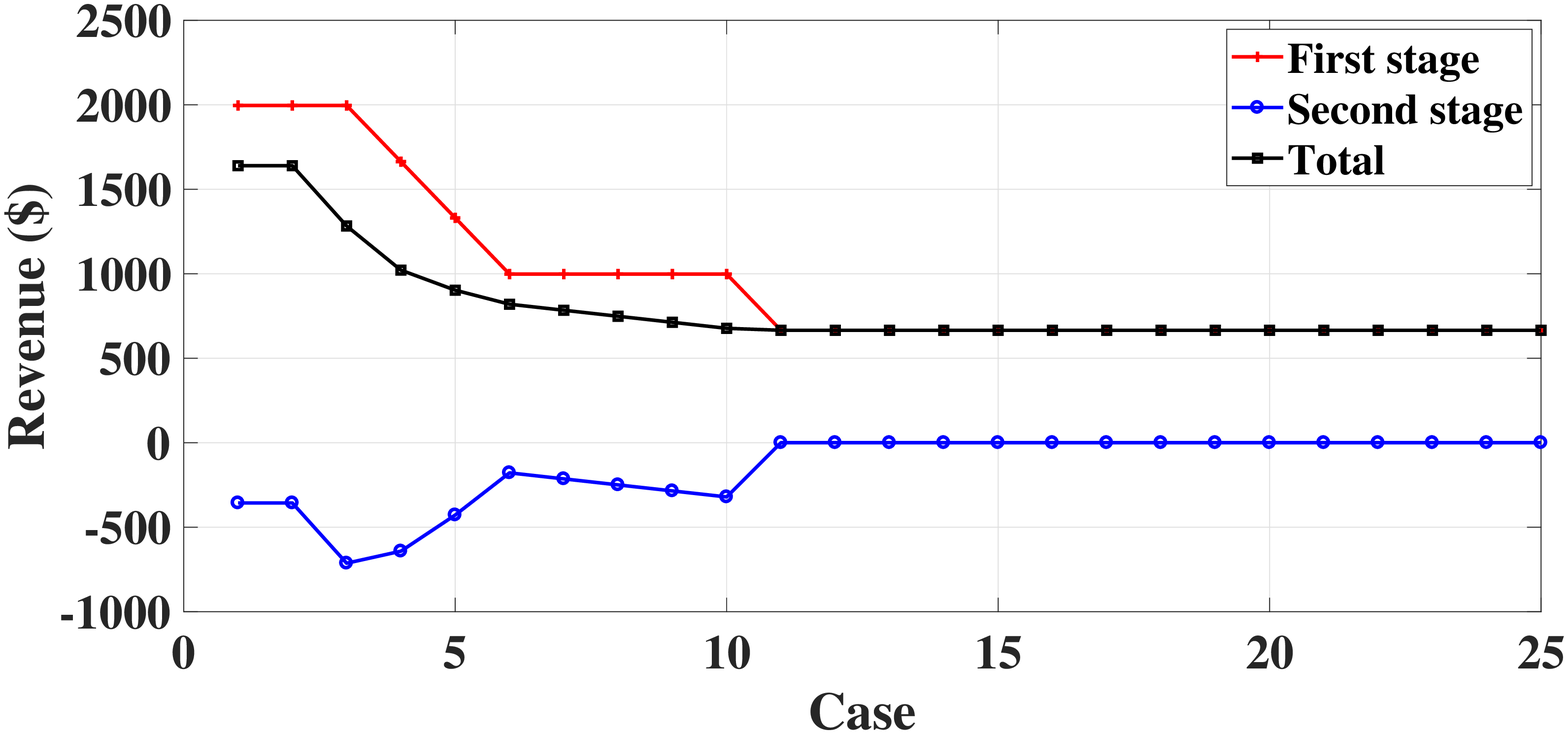}\label{fig.29.renewableaggregatorrevenuechange}
}

\subfigure[]
{
	\centering
	\includegraphics[width=0.93\columnwidth]{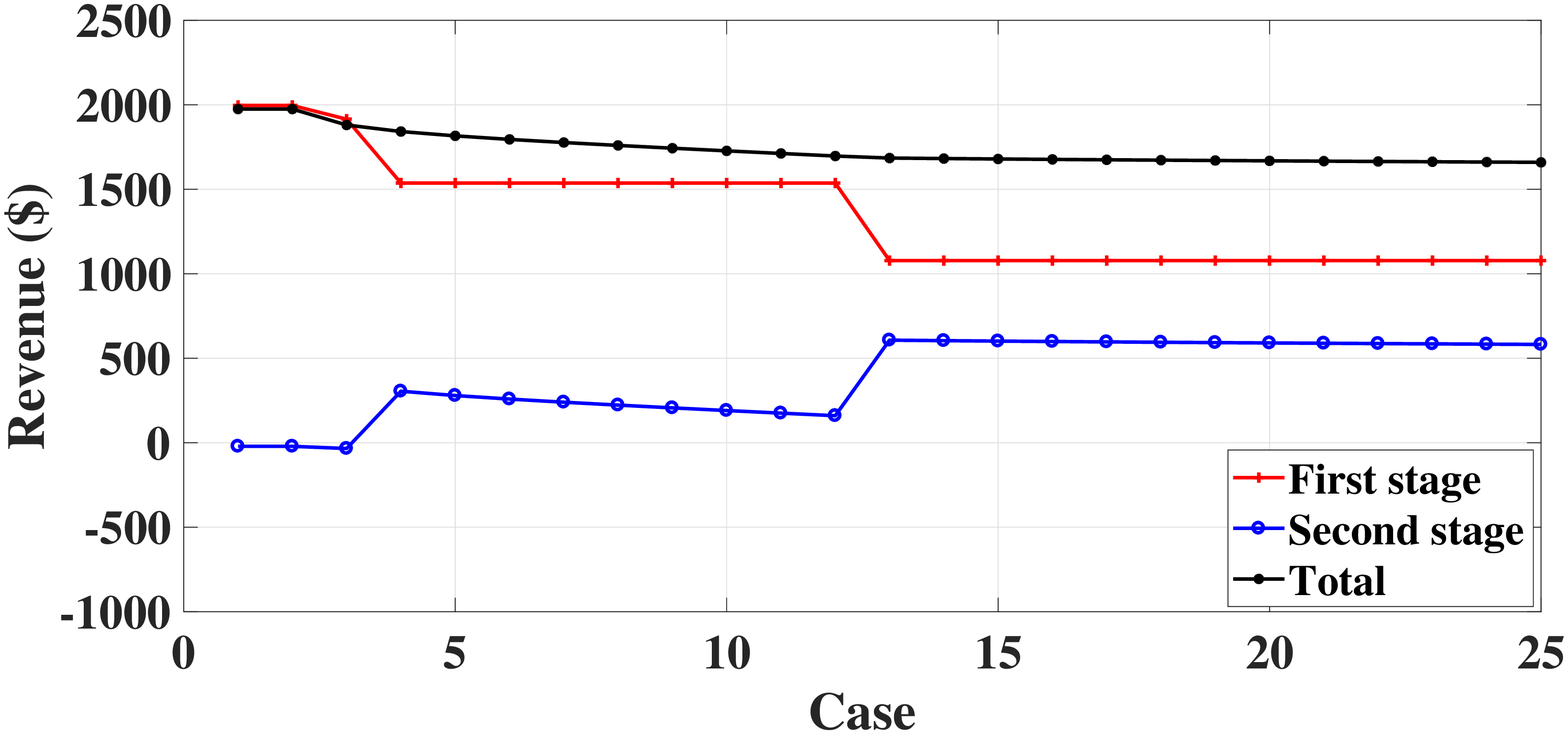}\label{fig.32.renewableaggregatorrevenuechange2}
}
	\caption{Changes in REAG's first-stage (day-ahead) revenue, expected second-stage (real-time) revenue, and total revenue with respect to changes in the real time prices under (a) one source of uncertainty; (b) multiple sources of uncertainties.}\label{Fig.33.renewableaggregatorrevenue}
\end{figure}

Fig. \ref{fig.29.renewableaggregatorrevenuechange} shows the changes in REAG's first-stage (day-ahead) revenue, expected second-stage (real-time) revenue, and total revenue with respect to changes in the real-time prices under one source of uncertainty. $25$ sensitivity cases are simulated. In each case, the base-case wholesale real-time market prices are multiplied by $i$, where $i$ varies from $1$ to $25$. When $i=1$, the REAG's second-stage (real-time) compensation cost is very low. Hence, its first-stage (day-ahead) revenue is high. Also, the REAG's second-stage (real-time) revenue is negative, which indicates the REAG buys power from the real-time market to compensate its power deficiency. Two factors affect the second stage revenues: 1) real-time prices; 2) amount of power deficiency that should be compensated in the real-time market. These two factors are negatively correlated with each other, which means when one factor increases the other factor decreases. The total effect of the two factors depends on the studied sensitivity case. For instance, when $i=3$, effect of real-time price on second-stage revenue is higher than that of power deficiency, which decreases the second-stage revenue. However, when $i$ increases, the effect of power deficiency grows. Hence, the second-stage revenue becomes zero after $i=10$. %However, the REAG's first-stage revenue and total revenue always decrease when the real-time prices increase.

Fig. \ref{fig.32.renewableaggregatorrevenuechange2} shows the changes in REAG's first-stage (day-ahead) revenue, expected second-stage (real-time) revenue, and total revenue with respect to changes in the real-time prices under multiple sources of uncertainties. To increase REAG's real-time compensation cost, REAG's real-time selling/purchasing prices are multiplied/divided by $i$, where $i$ varies from $1$ to $25$. When $i$ is small, the real-time compensation cost is low. Hence, the DSO schedules the REAG production at its mean value and covers the variations of inelastic load and REAG production by trading with the wholesale market. When $i$ increases, the real-time compensation cost becomes expensive. As a result, the DSO schedules the REAG production at a lower level to avoid trading with the wholesale market and compensate inelastic load variation by REAG production. This causes the REAG's expected second-stage (real-time) revenue to increase when $i$ becomes greater. %After $i=14$, the DSO schedules the minimum production for REAG, which causes the curves to become constant.

\section{Conclusion}
This paper proposes a two-stage stochastic programming DSO framework for coordination of DER aggregators to participate in the retail market as well as wholesale energy and regulation markets. Various kinds of DER aggregators are modeled in the proposed DSO framework. Case studies carried out on a small distribution network show key factors between the first-stage (day-ahead) and second-stage (real-time) LMPs. The REAG participates in day-ahead and real-time markets with uncertainties. Sensitivity analysis shows as the real-time price increases, the DSO schedules less power production to REAG as an uncertain market participant.

% conference papers do not normally have an appendix

% use section* for acknowledgment
%\section*{Acknowledgment}
%
%
%The authors would like to thank...

\bibliographystyle{IEEEtran}
\bibliography{References}

% that's all folks
\end{document}